\newcommand{\dm}{\delta m^2}
\newcommand{\be}{\begin{eqnarray}}
\newcommand{\ee}{\end{eqnarray}}
\newcommand{\bi}{\bibitem}
\newcommand{\nue}{\nu_e}
\newcommand{\num}{\nu_\mu}
\newcommand{\nut}{\nu_\tau}
\newcommand{\nus}{\nu_s}

\newcommand{\rar}{\rightarrow}
\newcommand{\lrar}{\leftrightarrow}

\documentclass{ws-procs9x6}

\begin{document}

\title{
Cosmological Constraints on Neutrino Masses and Mixings.
}

\author{A.~D. DOLGOV
}

\address{
INFN, sezione di Ferrara,
Via Paradiso, 12 - 44100 Ferrara,
Italy \\
and \\
ITEP, Bol. Cheremushkinskaya 25, Moscow 113259, Russia\\
E-mail: dolgov@fe.infn.it
}

\maketitle

\abstracts{
The bounds on neutrino masses and mixing that follows from the data on 
light element abundances, large scale structure formation, and angular
fluctuations of cosmic microwave background radiation are analyzed. 
The role of neutrino oscillations in BBN and the bounds on cosmological
lepton asymmetry are discussed.
}

\section{Introduction \label{s-intro}}

Neutrinos have the weakest interactions among all known elementary 
particles. They have also the smallest mass among all known massive 
particles. These two properties, on one hand, make it difficult to 
study neutrino properties directly, in particular, to measure their 
mass in laboratories.
On the other hand, the same properties make neutrinos very important
cosmologically and, at the present time, measuring neutrino masses looking
at the sky seems more promising than terrestrial experiments.
Significant cosmological role played by neutrinos arises from their large
number density. Neutrinos are the second most abundant particles in the 
universe, after photons in Cosmic Microwave Background Radiation (CMBR)
with the number density $n_\gamma \approx 410/{\rm cm}^3$. According to 
the standard cosmology, the universe is filled, in addition to CMBR, by 
the Cosmic Neutrino Background Radiation (C$\nu$BR) with the 
present-day number density:
\be
n^{(0)}_{\nu_a} = n^{(0)}_{\bar\nu_a} \approx 56/{\rm cm}^3
\label{n-nua}
\ee
for any neutrino flavor $a=e,\mu,\tau$. It is usually (but not always)
assumed that neutrinos are not degenerate (i.e. their chemical potentials 
are zero or negligibly small) and the number densities of neutrinos and 
antineutrinos are equal.

However, contrary to well observed CMBR, the existence of C$\nu$BR is 
only a theoretical prediction, and though practically nobody has any 
doubts of that, direct observation of C$\nu$BR is still missing and it 
seems that there is no chance for direct registration of cosmic neutrinos 
today and maybe even in the foreseeable future. Thus, one has to rely
on indirect methods studying features imprinted by cosmic neutrinos
on:
\begin{enumerate}
\item{}
formation and evolution of astronomical Large Scale Structure (LSS);
\item{}
angular fluctuations of CMBR;
\item{}
light element abundances created at Big Bang Nucleosynthesis (BBN);
\item{}
propagation of Ultra High Energy Cosmic Rays (UHECR).
\end{enumerate}

In what follows we will consider the first three subjects only. 
One can find a more detailed discussion of the issues presented below, 
as well as of the problem of interaction of UHECR with C$\nu$BR,
and hopefully a complete list of references in the review~\cite{dolgov-nu}.

\section{Early History of Neutrinos.\label{s-history} }

When the temperature of the cosmic plasma was above a few MeV, neutrinos
were in thermal equilibrium with the electromagnetic component of the 
plasma, i.e. with photons and $e^+e^-$-pairs. Electronic neutrinos
decoupled from $e^+e^-$-pairs when the temperature dropped below 
$T_d^{(e)}\approx 1.9$ MeV,
while $\num$ and $\nut$ decoupled a little earlier at 
$T_d^{(\mu)}\approx 3.1$ MeV.
At the moment of decoupling and later down to $T\approx m_e = 0.511$ MeV
temperatures of neutrino and electromagnetic components were equal,
$T_\nu=T_\gamma$. Below $T=m_e$ the annihilation of $e^+e^-$-pairs heats 
up photons and themselves, while leaves neutrino temperature intact. As a 
result the initial equilibrium ratio of neutrino-to-photon number densities
becomes diluted by the factor 4/11:
\be
\frac{n_\nu + n_{\bar \nu}}{ n_\gamma } =\frac {3}{4} \rar \frac{3}{11}
\label{dlt}
\ee
From this result the present-day neutrino number density (\ref{n-nua})
is obtained. Any additional energy release after $T_d^{(a)}$ which might
increase the photon number density would correspondingly diminish 
$n_\nu^{(0)}$.

Neutrino spectrum is close to the equilibrium form:
\be
f_\nu^{(eq)} = \left[ \exp (p/T_\nu -\xi) +1 \right]^{-1}
\label{f-nu}
\ee
where $p$ is the neutrino momentum, $T_\nu$ is the temperature, and
$\xi = \mu/T$ is dimensionless chemical potential; for adiabatic 
expansion $\xi$ remains constant. Usually chemical potentials of different
neutrino species are assumed to be negligible, at the level of baryonic
asymmetry, $10^{-9}$, though much larger values, even close to 1, are 
not excluded. Moreover, there exist theoretical models which predict a small
baryon asymmetry and simultaneously large lepton ones. Equilibrium with 
respect to $\bar\nu \nu$-annihilation enforces $\xi_\nu+\xi_{\bar\nu}=0$
but if neutrino charge asymmetry is generated at low temperatures this 
condition may be violated.
 
One more comment is in order: if neutrino mass is non-negligible 
in comparison with the temperature,
the spectrum (\ref{f-nu}) is non-equilibrium because the latter contains
$\exp (E/T)$ but not $\exp (p/T)$. At the present time 
$T_\nu \approx 0.714 T_\gamma \approx 1.68\cdot 10^{-4}$ eV. And if the 
neutrino mass is larger than this value the deviations from the usual
Fermi distribution may be significant. This must be true at least for
two out of three neutrinos because from the atmospheric neutrino
anomaly $\dm = (2-5)\cdot 10^{-3}$ eV$^2$ and from the solar neutrino
deficit and KAMLAND data $\dm = (6-7)\cdot 10^{-5}$ eV$^2$.

However, even at high temperatures ($T\gg m_\nu$) a deviation from 
equilibrium was non-negligible. Indeed,
the decoupling of neutrinos from $e^+e^-$-pairs is not instantaneous
and the annihilation $e^+e^- \rar \nu \bar\nu$ at $T\leq m_e$ would heat
the neutrino component of the plasma and distort its spectrum. According 
to analytical estimates of ref.~\cite{ad-fuk} the spectral distortion for
$\nue$ has the form:
\be
\frac{\delta f_{\nue}}{f_{\nue}} \approx 3\cdot 10^{-4}\,
\frac{E}{T}\,\left(\frac{11}{4}\,\frac{E}{T} -3\right) 
\label{nue-dstr}
\ee
Most accurate numerical solution of the kinetic equation that governs
non-equilibrium corrections to the neutrino spectrum was performed
in ref.~\cite{dhs0}. According to the calculations the excess of energy
density of $\nue$ and $\nu_{\mu,\tau}$ are respectively
$\delta \rho_{\nue} /\rho_\nu = 0.9\%$ and
$\delta \rho_{\num,\nut} /\rho_\nu = 0.4\%$. Together with the plasma
corrections~\cite{heckler} which diminish the energy density of the
electromagnetic component, the total relative rise of neutrino energy
density reaches approximately 4\%. This phenomenon has very little impact
on production of primordial $^4$He but can be observable in the shape of
the angular fluctuations of CMBR in the forthcoming Planck mission. If
observed, then together with BBN, it would present evidence of physical
processes which took place in the universe when she was about 1 sec old.
The corresponding red-shift is about $10^{10}$.
More details and references relevant to the subject of this section can be 
found in the review~\cite{dolgov-nu}.

\section{Gerstein-Zeldovich and LSS Bounds on Neutrino Mass \label{s-gz}}

Knowing the present-day number density of relic neutrinos one can easily
calculate their energy density and obtain an upper limit on their mass.
Such bound was derived in 1966 by Gerstein and Zeldovich~\cite{gz}
The result was re-derived 6 years later by Cowsic and Mc Lelland~\cite{c-ml}
but in their work the effect was overestimated by the factor 22/3.
In contemporary form the limits reads:
\be
\sum_a m_{\nu_a} \leq 94\,{\rm eV}\,\Omega h^2 
\label{gz}
\ee
where the sum is taken over all neutrino species, $a=e,\mu,\tau$;
$\Omega=\rho/\rho_c$ is the cosmological mass fraction of matter,
$\rho_c = 10.5\,h^2\,{\rm keV/cm}^3$, and $h$ is the dimensionless Hubble 
parameter, $h=H/100\,{\rm km/sec/Mpc} \approx 0.7$. 

According to the different and independent pieces of astronomical data  
$\Omega <0.3$ and correspondingly $\sum_a m_{\nu_a} < 14 $ eV. Since the 
data on neutrino oscillations~\cite{noon} show that the neutrino mass 
difference is much
smaller than eV, the mass of any neutrino flavor should be below 4.7 eV.

This limit can be further improved if one takes into account a possible
role that massive neutrinos might play in the process of formation of
the large scale structure of the universe. The point is that in neutrino
dominated universe all the structures on the scales smaller than the 
neutrino free streaming length, $l_\nu$, should be erased. The latter can be 
estimated as the distance that neutrinos could free-stream in the universe
before they became non-relativistic. The mass inside the free-streaming 
volume can be estimated as 
$M_\nu \approx 5\cdot 10^{17}\,M_\odot (1 {\rm eV} /m_\nu)^2$. 
If density perturbations are common for all particles (this is true for
adiabatic perturbations created by inflation), then neutrino out-stream 
would leave behind less power at small scales inhibiting structure formation
at these scales. The larger is $\Omega_\nu$ the larger is the effect. 
Moreover, the larger is the fraction of relativistic (hot) dark matter
the later structure formation begins. Hence observation of structures 
at large red-shifts allows to conclude that $\Omega_\nu < 0.1$ and
$m_\nu < 1.5$ eV~\cite{lss-mnu}. The analysis of the recent data 
from 2dF Galaxy Redshift Survey~\cite{2df} permitted to put the
limit $\sum_a m_{\nu_a}< 2.2$ eV or individual masses should be below 
0.73 eV. As argued in ref.~\cite{sdss} detailed
analysis of structure formation by the Sloan Digital Sky Survey would be
sensitive to neutrino mass at the level of (a few)$\times 0.1$ eV.   

This limits are based on certain assumptions about the form of the spectrum 
of perturbations, their character (adiabatic or isocurvature), and, what
is probably the safest one, about neutrino interactions. A question may
arise in this connection whether neutrinos can supply all the dark matter 
in the universe if we relax any or all these assumptions, introducing an 
arbitrary shape of the perturbation spectrum and/or new neutrino interactions
(of course inside the established limits). The answer would still be negative
because of the Tremain-Gunn limit~\cite{tremain}. This limit manifests
quantum mechanical Fermi exclusion principle at the kiloparsec scales and
demands that neutrinos must be heavier than roughly speaking 100 eV if
they form all dark matter in galaxies. So it seems that the only way out
is to make a crazy assumption that neutrinos are bosons and not fermions
which in light of the discussed today search of CPT violation maybe looks 
not so crazy because CPT-theorem is heavily based on the standard relation
between spin and statistics.

\section{Neutrino Mass and CMBR \label{s-cmbr}}

The impact of neutrinos on the shape of angular spectrum of CMBR temperature
is based on the following two effects (for details see e.g. the
recent review~\cite{cmbr}). First, a change of the energy density 
of relativistic matter (neutrinos) would change the cosmological expansion 
regime and this in turn would change the physical size of the horizon at 
recombination. It would shift the positions of the acoustic peaks in
the temperature fluctuations. This effect is relatively weak. More 
important is another one which changes the heights of the peaks because
an admixture of relativistic matter gives rise to an amplification of 
acoustic oscillations since relativistic matter creates gravitational
force which varies with time creating resonance amplification of the
acoustic oscillations. These phenomena permit to measure neutrino mass and
the number of neutrino families (or in other words, the energy density of
neutrinos) in the recombination epoch. 

The recent measurements of the angular fluctuations of CMBR by 
WMAP~\cite{wmap}, together with the analysis of LSS by 2dF,
permitted to impose surprisingly strong upper limit on neutrino mass
\be
\sum_a m_{\nu_a} < 0.69\,{\rm eV},
\label{wmap}
\ee 
that is for mass degenerate neutrinos $m_\nu < 0.23$ eV at 95\% CL;
see also the discussion of this result and of the role of primers in 
ref.~\cite{elgaroy03}.

Immediately after publication of the first WMAP data, there appeared 
several papers~\cite{n-nu-cmbr} where the number of neutrino families were 
evaluated. Their results are, roughly speaking, $N_\nu = 1-7$ depending
upon the analysis and priors. Anyhow $N_\nu >0$ and thus CMBR confirms,
independently on BBN, that C$\nu$BR (or some other relativistic background)
indeed exists.

The accuracy of the data is not good enough to compete with the
determination of $N_\nu$ from BBN (see sec.~\ref{s-bbn}) to say nothing
about a registration of 4\% addition to the neutrino energy discussed in
sec.~\ref{s-history}. However the forthcoming Planck mission may be sensitive
to this effect and overtake BBN in determination of $N_\nu$. One should keep
in mind that the impact of neutrino energy on BBN and CMBR may be different
depending upon the form of the spectral distortion of electronic neutrinos
$f_{\nue} (E)$. The effect, which was calculated for massless neutrinos,
also depends upon the value of neutrino mass. This may be essential at the
recombination epoch.

\section{Neutrinos and BBN \label{s-bbn}}

Abundances of light elements ($^2$H, $^3$He, $^4$He, and $^7$Li) produced
at BBN at $T = 1-0.07$ MeV (t=1-200 sec) depend upon the following 
quantities:
\begin{enumerate}
\item{}
Number density of baryons, $\eta_{10} = 10^{10} n_B/n_\gamma$. In the previous
century the value of this parameter was determined from BBN itself through
comparison of the predicted deuterium abundance with observations. Now after
measurements of CMBR angular fluctuations by MAXIMA, BOOMERANG and 
DASI~\cite{eta} and confirmed by WMAP~\cite{wmap} this parameter is 
independently fixed at $\eta_{10} = 6\pm 0.3$.
\item{}
Weak interaction rate which is expressed through the neutron life-time,
now well established, $885.7\pm 0.8$ sec~\cite{pdg}. 
\item{}
Cosmological energy density at the period of BBN. The latter is usually
parametrized as the effective number of additional neutrino species
$\Delta N_\nu = N_\nu -3$. This type of parametrization is flawless for
relativistic energy, while for another form of energy (e.g. non-relativistic
or vacuum-like) its effect on the production of different light elements may 
deviate from that induced by neutrinos.  
\item{}
Neutrino degeneracy, i.e. a possible non-vanishing values of neutrino
chemical potentials $\xi_a$. While for $\num$ and $\nut$ non-zero
$\xi_{\mu,\tau}$ are equivalent to an increase of $N_\nu$, degeneracy
of electronic neutrinos has a much stronger impact on BBN because the 
frozen neutron-to-proton ratio is exponentially sensitive to the magnitude
(and sign) of chemical potential of electronic neutrinos,
$n_n/n_p \sim \exp (-\xi_{\nue})$.
\item{}
Energy spectrum of $\nue$. If the spectrum is distorted then the equilibrium
of reactions $n\nue\lrar pe^-$ and $ne^+\lrar p\bar\nue$ is shifted and
could result either in an increase or decrease of $n_n/n_p$ depending upon
the form of distortion.
\end{enumerate}

The upper bound on the effective number of neutrino families, found from 
BBN, had a rather strong time evolution and it changed from 
$\Delta N_\nu = 3-5$ in the original papers~\cite{n-nu}
down to 0.3 and even to 0.1 
in subsequent literature. The existing data on the light element abundances 
is still controversial~\cite{ad-taup} and possibly the situation described
in ref.~\cite{lisi99} still remains true, which is roughly the following
$n_\nu = 3\pm 0.5$. 

The most accurate upper bound on chemical potential of a single neutrino
species under the conservative assumptions that $\Delta N <1$ and other 
chemical potentials are vanishingly small was obtained in 
ref.~\cite{kohri97} and reads: $|\xi_{\num,\nut}| <1.5$ and
$|\xi_{\nue}| < 0.1$. This bound would be relaxed if a conspiracy between
different chemical potentials is allowed such that the effect of a large
$|\xi_{\num,\nut}|$ is compensated by $|\xi_{\nue}|$ (or vice versa). The 
best limit is presented in ref.~\cite{hansen01} where the combined inputs
from BBN and CMBR have been used: $|\xi_{\num,\nut}|< 2.6 $ and
$|\xi_{\nue}| < 0.2$. 

The bounds quoted above are valid if neutrinos are not mixed. Otherwise,
charge asymmetry of a certain neutrino flavor would be redistributed between
all neutrino species and the bound would be determined by the most sensitive
asymmetry of $\nue$. This problem was analyzed in ref.~\cite{dhpprs} 
numerically and analytically in ref.~\cite{dolgov-nu}, where
it was shown that for Large Mixing Angle solution of the solar neutrino 
deficit the transformation is quite efficient and equilibration of all
chemical potentials by oscillations is achieved. This leads to the common
bound for all neutrino chemical potentials
\be
|\xi_a | < 0.07
\label{xi-a}
\ee
Similar investigation both analytical and numerical was also performed 
in the papers~\cite{aa}.

If a new sterile neutrino (or several sterile species) mixed with active 
ones exists, the impact of neutrino oscillation on BBN would be more 
interesting than in the case of active-active mixing. The 
$\nus\lrar\nu_a$-transformations would excite additional neutrino species
leading to an increase of the effective number of neutrinos,
$\Delta N >0$~\cite{dolgov81,barbieri90}, could distort the 
spectrum of $\nue$~\cite{barbieri90,kirilova97} and could generate a large
lepton asymmetry in the sector of active neutrinos
by MSW-resonance~\cite{l-asym}. (More references and
discussion can be found in~\cite{dolgov-nu}.)

In the non-resonance case, i.e. for $m_{\nus}>m_{\nu_a}$ it is rather 
easy to
estimate the production rate of sterile neutrinos in the early universe
through oscillations and to obtain the following bounds on the oscillation
parameters~\cite{ad-itep}:
\be
(\dm_{\nue\nus}/{\rm eV}^2) \sin^4 2\theta_{vac}^{\nue\nus} =
3.16\cdot 10^{-5}\,\left[\log \left( 1-\Delta N_\nu\right)\right]^2
\label{dmess2}\\
(\dm_{\num\nus}/{\rm eV}^2) \sin^4 2\theta_{vac}^{\num\nus} =
1.74\cdot 10^{-5}\, \left[\log \left( 1-\Delta N_\nu\right)\right]^2
\label{dmmuss2}
\ee
In this result a possible deficit of $\nue$ created by the transformation
of the latter into $\nus$, when refilling of $\nue$ by $e^+e^-$-annihilation
is already weak, is not taken into account. This effect would strengthen the 
bound. 

Numerical solution of kinetic equations governing neutrino oscillations in 
the early universe was performed in ref.~\cite{enqvist92} under assumption
of kinetic equilibrium, so the neutrinos are described by a single momentum
state with the thermally average value of the energy $E=3.15 T$. For the 
non-resonant case the obtained results are somewhat stronger than those
presented above, (\ref{dmess2},\ref{dmmuss2}). For the resonant case it 
is questionable if thermal averaging is a good approximation because the
position of resonance depends upon the neutrino momentum. One more 
complication is that now is known that all active neutrinos are strongly
mixed and their mutual transformation should be taken into account together
with $(\nu_a-\nus)$-oscillations~\cite{dv}. The active neutrino mixing
noticeably changes the previously obtained cosmological limits for mixing
with $\nus$. Moreover, the solution of momentum dependent kinetic equations
shows that kinetic equilibrium is strongly broken (at least for some values
of the oscillation parameters) and, in particular, the 
spectrum of $\nue$ is distorted leading to a shift of the $n/p$-ratio for
interesting values of mass difference between $\nus$ and active neutrinos. 
The calculations are complicated by the appearance of one, two or even three 
resonances if sterile neutrino is lighter than one, two or all three active 
ones. Anyhow in the resonance case the cosmological bounds on 
the mixing between
sterile and active neutrinos are considerably stronger than those in 
non-resonance case. Thus if a large mixing to $\nus$ is discovered it would
mean that the lepton asymmetry of the universe is non-negligible~\cite{dv},
because the latter might ``cure'' the effect of $\nu_a-\nus$ oscillations on 
BBN.

\section{Conclusion \label{s-concl}}

Thus we see that cosmology is becoming sensitive to the values of neutrino
masses approaching $\sqrt{\dm}$. So one may hope that neutrino will be
the first particle whose mass will be measured by astronomers by the 
combined data from CMBR and LSS.

The number of additional neutrinos at BBN is limited by 0.5 (though 1 is
probably still not excluded) with a possibility to improve this limit down 
to 0.1.

The observed in experiment strong mixing between active neutrinos allows
cosmological lepton asymmetry to be relatively low, smaller than 0.1. This
excludes, in particular, cosmological models where large chemical potential 
of neutrinos might be essential for large scale structure formation.

A possible mixing between active and sterile neutrinos is restricted by
BBN at much stronger level than by direct experiment.

Planck mission may be sensitive to additional contribution to neutrino
energy density at per cent level and thus will be able to trace physical
processes in the universe at red shift of $10^{10}$.

I am grateful to F. Villante for critical comments.


\begin{thebibliography}{0}

\bibitem{dolgov-nu}
A.D. Dolgov, {\it Phys. Repts.} {\bf 370}, 333 (2002).

\bibitem{ad-fuk}
A.D.~Dolgov and M.~Fukugita, {\it JETP Lett.} {\bf 56} (1992) 123
[{\it Pisma Zh.\ Eksp.\ Teor.\ Fiz. }  {\bf 56} (1992) 129];\\
A.D.~Dolgov and M.~Fukugita, {\it Phys. Rev.} {\bf D46}, 5378 (1992).

\bibitem{dhs0} 
A.D.~Dolgov, S.H.~Hansen and D.V.~Semikoz, {\it Nucl. Phys.} {\bf B503} 
(1997) 426; {\it Nucl. Phys.} {\bf B543} (1999) 269.


\bibitem{heckler} 
A.F.~Heckler, {\it Phys. Rev.} {\bf D49 }, 611 (1994);\\
R.E.~Lopez, S.~Dodelson, A.~Heckler and M.S.~Turner,
{\it Phys. Rev. Lett.} {\bf 82}, 3952 (1999).

\bibitem{gz}
S.S. Gerstein and Ya.B. Zeldovich, {\it  Pis'ma ZhETF}, {\bf 4}, 174 (1966)
[English translation {\it JETP Letters} {\bf 4}, 120 (1966)].

\bibitem{c-ml}
R.~Cowsik and J.~McClelland,
{\it Phys. Rev. Lett.}  {\bf 29}, 669 (1972).

\bibitem{noon}
See e.g. the talks by E. Lisi, T. Schwetz, A.Y. Smirnov at this Conference.

\bibitem{lss-mnu}
R.A.C. Croft, W. Hu, and R. Dav{\'e}, {\it Phys. Rev. Lett.} {\bf 83},
1092 (1999);\\
W.~Hu, D.J.~Eisenstein and M.~Tegmark, {\it Phys. Rev. Lett.} {\bf 80}, 
5255 (1998);\\
M. Fukugita, G.-C. Liu, N. Sugiyama, {\it Phys. Rev. Lett.} {\bf 84},
1082 (2000);\\
A.R.~Cooray, {\it Astron. Astrophys.} {\bf 348}, 31 (1999);\\ 
E.~Gawiser, Proceedings of PASCOS99 Conference, Lake Tahoe, CA 1999,
astro-ph/0005475.

\bibitem{sdss}
W.~Hu, D.J.~Eisenstein, M.~Tegmark and M.J.~White,
{\it Phys. Rev.}  {\bf D59}, 023512 (1999).

\bibitem{2df}
O. Elgaroy et al, {\it Phys. Rev. Lett.} {\bf 89}, 061301 (2002).

\bibitem{tremain}
S.~Tremaine and J.E.~Gunn,  {\it Phys. Rev. Lett.} {\bf 42}, 407
(1979).

\bibitem{cmbr}
W. Hu and S. Dodelson, {\it Annu. Rev. Astron. and Astrophys.} 
{\bf 40}, 171 (2002).

\bibitem{wmap}
D.N. Spergel et al, astro-ph/0302209.

\bibitem{elgaroy03}
O. Elgaroy and O. Lahav, {\it JCAP} {\bf 0304}, 004 (2003).

\bibitem{n-nu-cmbr}
P. Crotty, J. Lesgourgues and S. Pastor, astro-ph/0302337;\\
S. Hannestad, astro-ph/0303076;\\
E. Pierpaoli, astro-ph/0302465;\\
V. Barger, J.P. Kneller, H.-S. Lee, D. Marfatia and G. Steigman,
hep-ph/0305075.

\bibitem{eta}
A.T. Lee et al, {\it Astrophys. J.} {\bf 561}, L1 (2001);\\
C.B. Netterfield et al, {\it Astrophys. J.} {\bf 571} (2002) 604;\\
N.W. Halverson et al, {\it Astrophys. J.} {\bf 568} (2002) 38.

\bibitem{pdg}
K. Hagiwara et al, {\it Phys. Rev.} {\bf D66},010001 (2002).

\bibitem{n-nu}
V.~F.~Shvartsman, {\it Pisma Zh. Eksp. Teor. Fiz.} {\bf 9}, 315 (1969)
[{\it JETP Lett.}  {\bf 9}, 184 (1969)];\\
G.~Steigman, D.N.~Schramm and J.R.~Gunn,
{\it Phys. Lett.} {\bf B66}, 202 (1977).

\bibitem{ad-taup}
A.D. Dolgov, {\it Nucl. Phys., Proc. Suppl.} {\bf B110}, 137 (2002).

\bibitem{lisi99}
E.~Lisi, S.~Sarkar and F.L.~Villante, {\it Phys. Rev.} {\bf D59}, 
123520 (1999);\\
G.~Fiorentini, E.~Lisi, S.~Sarkar and F.L.~Villante, {\it Phys. Rev.} 
{\bf D58}, 063506 (1998).

\bibitem{kohri97}
K.~Kohri, M.~Kawasaki and K.~Sato, {\it Astrophys. J.}  {\bf 490}, 72 (1997).

\bibitem{hansen01}
S.H.~Hansen, G.~Mangano, A.~Melchiorri, G.~Miele and O.~Pisanti,
{\it Phys. Rev.} {\bf D65}, 023511 (2002).

\bibitem{dhpprs}
A.D.~Dolgov, S.H.~Hansen, S.~Pastor, S.T.~Petcov, G.G.~Raffelt and
D.V.~Semikoz, {\it Nucl.Phys.} {\bf B632}  363 (2002).

\bibitem{aa}
C.Lunardini and A.Yu.Smirnov, {\it Phys. Rev.} {\bf D64}, 073006 (2001);\\
Y.Y.Y. Wong, {\it Phys. Rev.} {\bf D66} 025015 (2002);\\
K.N. Abazajian, J.F. Beacom and N.F. Bell, {\it Phys.Rev.} {\bf D66} 
013008 (2002).

\bibitem{dolgov81}
A.D.Dolgov, {\it Yad. Fiz.}  {\bf 33}, 1309 (1981); [English translation: 
{\it Sov. J. Nucl. Phys.} {\bf 33}, 700 (1981).

\bibitem{barbieri90}
R.~Barbieri and A.D.~Dolgov, {\it Phys. Lett.} {\bf B237}, 440 (1990);\\
K.~Kainulainen, {\it Phys. Lett.} {\bf B244}, 191 (1990).

\bibitem{kirilova97}
D.P. Kirilova and M.V. Chizhov, {\it Phys. Lett.} {\bf B393}, 375 (1997).

\bibitem{l-asym}
R.~Foot, M.J.~Thomson and R.R.~Volkas,
{\it Phys. Rev.} {\bf D53}, 5349 (1996).

\bibitem{ad-itep}
A.D. Dolgov, {\it Surveys in High Energy Physics}, {\bf 17}, 91 (2002) 
(lectures presented at ITEP Winter School, February, 2002).

\bibitem{enqvist92}
K.~Enqvist, K.~Kainulainen and M.~Thomson,
{\it Nucl. Phys.} {\bf B373}, 498 (1992).

\bibitem{dv}
A.D. Dolgov and F.L. Villante, work in progress.

\end{thebibliography}
\end{document}